# Geometry-Consistent Discretization of Local Correlation Equations for the Radon Transform in Clinical CT

**Xuanqing Mou*, Jiayu Duan and Yang Mou**
Institute of Image Processing and Pattern Recognition, School of Information and Communications Engineering, Xi'an Jiaotong University, Xi'an, People's Republic of China
* Author to whom any correspondence should be addressed.
E-mail: xqmou@mail.xjtu.edu.cn

## Abstract

In this work, we extend the local correlation equation (LCE) framework of the 2D Radon transform to clinical computed tomography (CT) geometry and propose a set of geometry-consistent discretization schemes. Building upon the dual-center-of-rotation formulation, the 1st and 2nd LCEs are reformulated under circular fan-beam geometry and discretized using finite difference methods aligned with clinical acquisition configurations. The numerical validity of the proposed discretized circular LCEs (cLCEs) is evaluated using projection data generated from clinical CT datasets. Quantitative analysis based on residual error, mean absolute error (MAE), and mean percentage error (MPE) demonstrates strong consistency between both sides of the equations, with the 2nd cLCE exhibiting the highest accuracy. Furthermore, the proposed cLCE framework is applied to sparse-view sinogram interpolation. By refining interpolated projections using a gradient-descent-based scheme without additional regularization, improved reconstruction quality is achieved. Experimental results show that the cLCE-refined projections significantly enhance image quality compared to sparse-view reconstruction and can outperform clinical complete view (CCV) reconstruction under certain conditions. These results demonstrate that the proposed cLCE discretization provides an effective tool for exploiting projection redundancy in clinical CT and has strong potential for sparse-view reconstruction applications.

## 1. Introduction

In our previous work (Mou and Duan 2025), we proposed a universal $n$th order partial differential equation (PDE) of the 2D Radon transform to characterize the intrinsic local correlation among neighboring projections. By introducing a dual-center-of-rotation geometry, the proposed local correlation equation (LCE) provides an object-independent relationship between the angular and distance derivatives of the 2D Radon transform. This formulation reveals the redundancy of projection data and offers a theoretical foundation for recovering complete-view sinograms from sparse-view acquisitions.

The original study focused on the theoretical derivation of the LCE and its preliminary discretization under idealized circular (cLCE) and stationary linear (sLCE) scanning trajectories. In particular, the discretization of the cLCE was primarily developed based on a unit-circle model, which is suitable for theoretical analysis but presents limitations for practical implementation. Specifically, this discretization leads to uneven sampling of certain derivative components (e.g., the parallel-beam-related $\frac{\partial R}{\partial t}$), and is therefore not directly compatible with projection data acquired under clinical CT geometry.

Clinical CT systems are governed by specific geometric configurations, including source–detector distances, fan-beam divergence, and discrete detector sampling (Kohl 2005). These factors necessitate a geometry-consistent discretization of the LCE for practical implementation. In this work, we extend the LCE framework to clinical CT geometry and propose a set of novel discretization schemes. Building upon the dual-center-of-rotation geometry introduced in the previous work, we first revisit the LCE formulation. We then incorporate the physical constraints of clinical CT systems to derive discretization schemes that are directly applicable to real-world data acquisition. In contrast to the previous work, the present study adopts a discretization protocol that is fully consistent with clinical CT geometry. This approach enables uniform sampling of the projection data and allows the LCE framework to be directly applied to clinically acquired measurements. Furthermore, the sinogram interpolation experiments are revisited using the proposed discretization scheme, demonstrating improved performance in recovering complete-view sinograms from sparse-view data.

## 2. LCE and dual-center-of-rotation geometry

The 2D Radon transform describes the line integral of an object along a set of parameterized lines defined by angular and distance variables. In conventional formulations, the relationship between angular and distance derivatives of the Radon transform is coupled with the object function (Natterer 2001) or depends on specific trajectory conditions (Patch 2002), which limits its applicability for constructing universal correlation models.

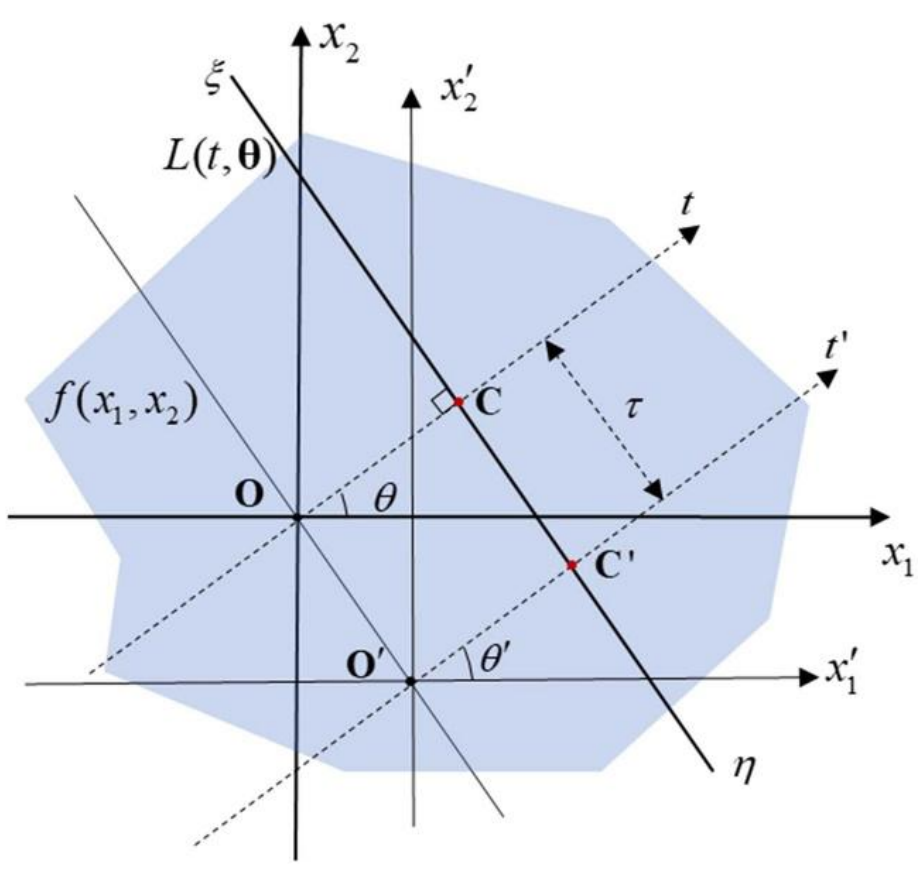

Figure 1. Illustration of the dual-center-of-rotation geometry (Mou and Duan 2025)

To address this limitation, we introduced a dual-center-of-rotation geometry (Figure 1), in which the same projection line is described with respect to two distinct coordinate systems (Mou and Duan 2025). By analyzing the angular derivatives evaluated at these two rotation centers, an object-independent relationship can be established. Specifically, the 1st order LCE expresses the distance derivative of the 2D Radon transform as the difference between angular derivatives evaluated at the two centers (Equation (1), where R denotes the Radon transform along the integration line $L(t,\theta)$). This formulation eliminates dependence on the imaging object and reveals an intrinsic local correlation within the sinogram.

$$\frac{\partial R}{\partial t} = \frac{1}{\tau}\left(\frac{\partial R}{\partial \theta'} - \frac{\partial R}{\partial \theta}\right) \tag{1}$$

The 1st order LCE can be extended to higher-order PDEs through successive differentiation. The resulting equations exhibit a structured form in which the coefficients follow a binomial expansion pattern. Equation (2) presents the 2nd order LCE derived from Equation (1). These higher-order LCEs further characterize the redundancy of the Radon transform and provide additional constraints for sinogram reconstruction and interpolation.

$$\frac{\partial^2 R}{\partial t^2} = \frac{1}{\tau^2}\frac{\partial^2 R}{\partial \theta'^2} - \frac{2}{\tau}\frac{\partial^2 R}{\partial t \partial \theta} - \frac{1}{\tau^2}\frac{\partial^2 R}{\partial \theta^2} \tag{2}$$

The LCE framework is general and can be adapted to different CT acquisition geometries by appropriately defining the dual centers. In particular, when one of the rotation centers is placed at the X-ray source, the LCE naturally becomes applicable to fan-beam geometries, forming the basis for the circular LCE (cLCE).

## 3. cLCE based on clinical CT geometry and its discretization

### 3.1 cLCE based on clinical CT geometry

Clinical CT systems typically adopt a circular scanning trajectory with an equiangular fan-beam geometry. The X-ray source rotates around the object along a circular path, while the detector array samples the attenuated projections at discrete angular and spatial intervals (Figure 2).

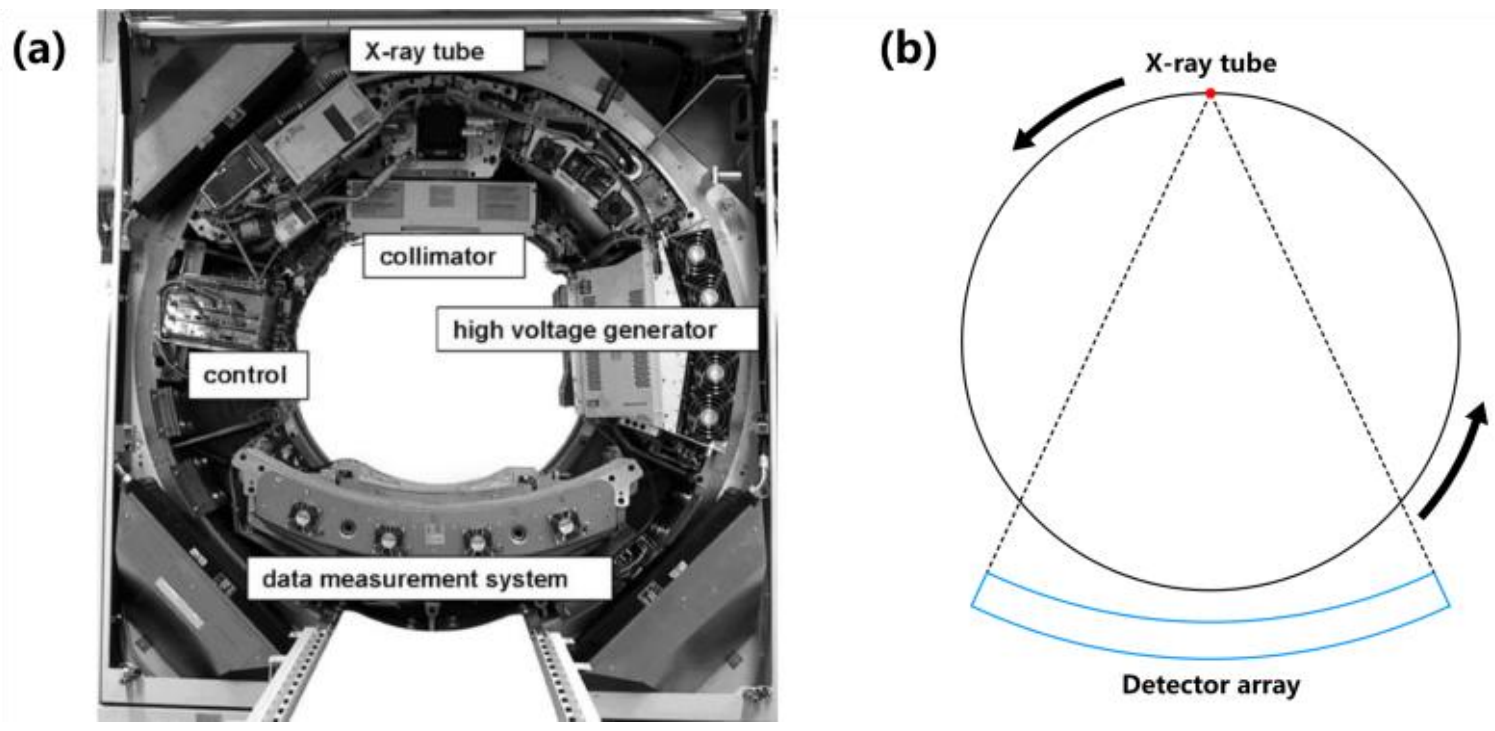

Figure 2. (a) Basic components of a modern 3rd-generation CT system (Kohl 2005). (b) Schematic illustrating the motion of the X-ray source and detector array

In this configuration, all projection rays corresponding to a given view originate from a common focal spot and intersect the detector array. The projection data are therefore naturally parameterized by the source angle and detector index, rather than the continuous Radon variables.

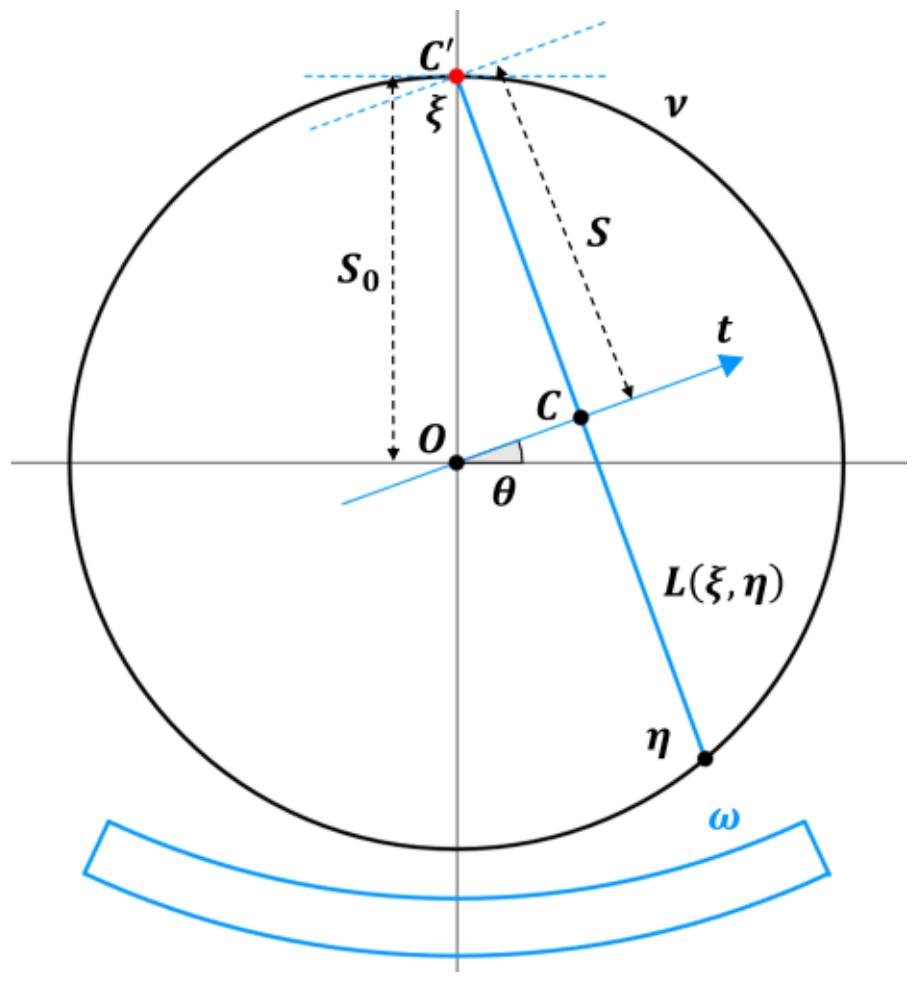

Figure 3. Dual-center-of-rotation geometry for circular scanning trajectory in clinical CT

As shown in Figure 3, the X-ray source position is denoted by $\xi$, the equiangular detector coordinate by $\omega$, and the integration line by $L(\xi,\eta)$. To adapt the LCE to clinical CT geometry, the dual-center-of-rotation structure is reinterpreted such that one rotation center (center $C'$) coincides with the X-ray source position $\xi$, while another rotation center (center $C$) is defined as the intersection point between the integration line $L(\xi,\eta)$ and the axis $t$.

Under this configuration, the 1st and 2nd order LCE can be reformulated in terms of source angle and detector coordinates, yielding the circular LCE (cLCE):

$$\frac{\partial R}{\partial t} = \frac{1}{s}\left(\frac{\partial R}{\partial \omega} - \frac{\partial R}{\partial \theta}\right) \tag{3}$$

$$\frac{\partial^2 R}{\partial t^2} = \frac{1}{s^2}\frac{\partial^2 R}{\partial \omega^2} - \frac{2}{s}\frac{\partial^2 R}{\partial \theta \partial t} - \frac{1}{s^2}\frac{\partial^2 R}{\partial \theta^2} \tag{4}$$

When the rotation center $C$ is shifted to $\eta$, an alternative form of the cLCE can be derived:

$$\frac{\partial R}{\partial t} = \frac{1}{s}\left(\frac{\partial R}{\partial \omega} - \frac{\partial R}{\partial \nu}\right) \tag{5}$$

$$\frac{\partial^2 R}{\partial t^2} = \frac{1}{s^2}\frac{\partial^2 R}{\partial \omega^2} - \frac{2}{s}\frac{\partial^2 R}{\partial \omega \partial \nu} + \frac{1}{s^2}\frac{\partial^2 R}{\partial \nu^2} \tag{6}$$

In Equation (5) and (6), the $\frac{\partial R}{\partial \nu}$ term corresponds to the angular derivative evaluated at the source position $\eta$, with $\nu$ serving as the detector coordinate, which is opposite in direction to the $\frac{\partial R}{\partial \omega}$ term. Although the rotation center $C$ can be arbitrarily defined along the integration line $L(\xi,\eta)$, we adopt the configuration in Equation (3) and (5) due to their suitability for discretization and numerical implementation.

### 3.2 Discretization of cLCE

In this section, we employ the finite difference method (FDM) to discretize the proposed cLCE. The discretization scheme based on clinical CT geometry is illustrated in Figure 4. The circular scanning trajectory is uniformly divided into M views, with $\Delta\omega$ denoting the circumferential angle interval between adjacent views. The equiangular detector array is discretized into N elements, with $\Delta\omega$ serving as the angle interval between adjacent rays. Let $R(m,n)$ denote the projection corresponding to the m-th view ($m \in [0,\cdots,M-1]$) and n-th detector ($n \in [0,\cdots,N-1]$). The parameter $\theta$ represents the angle between the current ray and the central ray, and $L$ denotes the intersection length of current ray and circular trajectory.

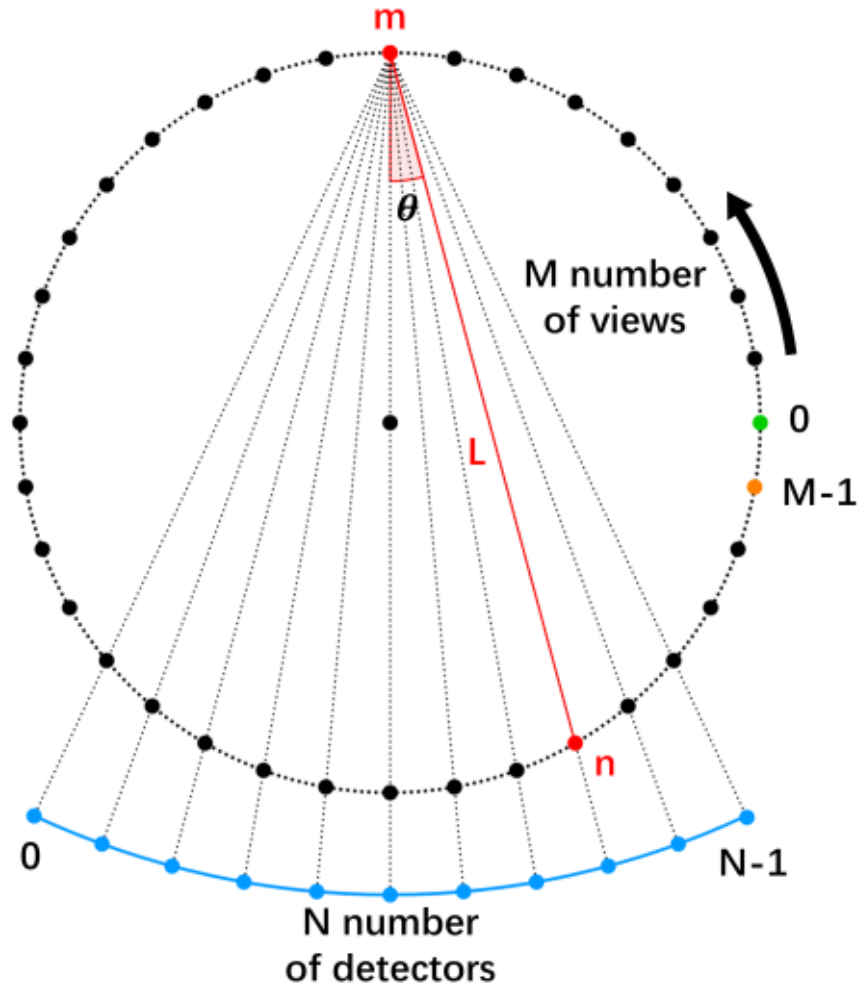

Figure 4. Indexing scheme for view and detector in cLCE discretization

The discretization of five representative 1st order cLCE, denoted as cLCE_i ($i = 1,\cdots,5$), is illustrated in Figure 5. Based on Equation (5), the discretized form of cLCE_1 can be obtained using the central difference method:

$$
\begin{aligned}
\frac{\Delta R}{\Delta t} &= \frac{R_{m-1,n+2} - R_{m+1,n-2}}{2\Delta\omega L} \\
\frac{\Delta R}{\Delta\omega} &= \frac{R_{m,n+1} - R_{m,n-1}}{2\Delta\omega} \\
\frac{\Delta R}{\Delta\nu} &= \frac{R_{m+1,n-1} - R_{m-1,n+1}}{2\Delta\omega} \\
s &= L
\end{aligned}
\tag{7}
$$

Substituting the finite difference approximation into Equation (5), cLCE_1 can be expressed as:

**cLCE_1:**

$$R_{m-1,n+2} - R_{m+1,n-2} = R_{m,n+1} - R_{m,n-1} - (R_{m+1,n-1} - R_{m-1,n+1}) \tag{8}$$

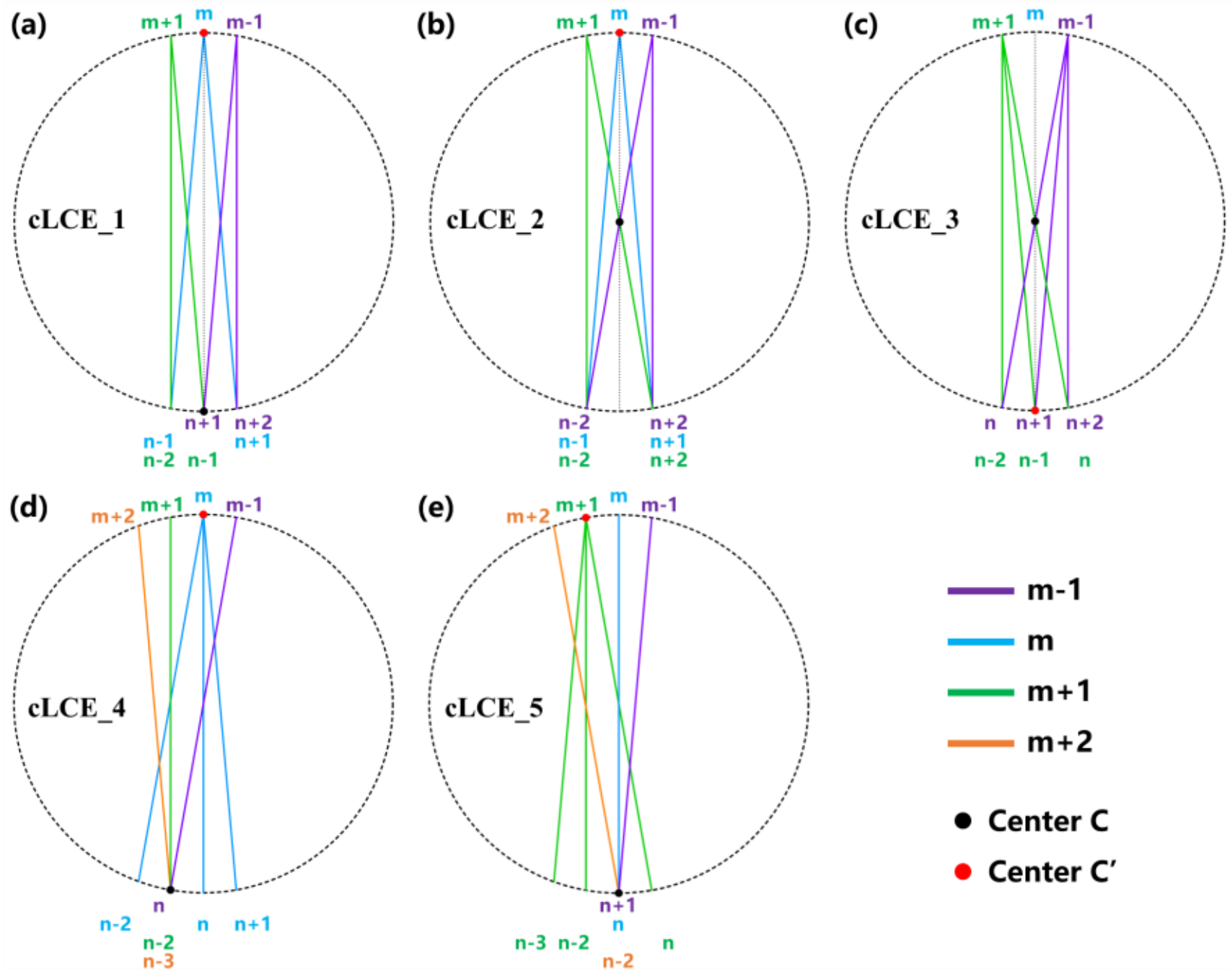

Figure 5. Discretization patterns for 1st order cLCEs

Using the same discretization strategy, the remaining 1st order cLCEs (cLCE_2 ~ cLCE_5) can be derived:

**cLCE_2:**

$$R_{m-1,n+2} - R_{m+1,n-2} = 2(R_{m,n+1} - R_{m,n-1}) - (R_{m+1,n} - R_{m-1,n}) \tag{9}$$

**cLCE_3:**

$$R_{m-1,n+2} - R_{m+1,n-2} = 2(R_{m-1,n+1} - R_{m+1,n-1}) - \left(R_{m-1,n} - R_{m+1,n}\right) \tag{10}$$

**cLCE_4:**

$$R_{m,n} - R_{m+1,n-2} + R_{m-1,n} - R_{m,n-2} = \frac{2}{3}\left(R_{m,n+1} - R_{m,n-2}\right) - \frac{2}{3}(R_{m+2,n-3} - R_{m-1,n}) \tag{11}$$

**cLCE_5:**

$$R_{m+1,n} - R_{m+2,n-2} + R_{m,n} - R_{m+1,n-2} = \frac{2}{3}(R_{m+1,n} - R_{m+1,n-3}) - \frac{2}{3}(R_{m+2,n-2} - R_{m-1,n+1}) \tag{12}$$

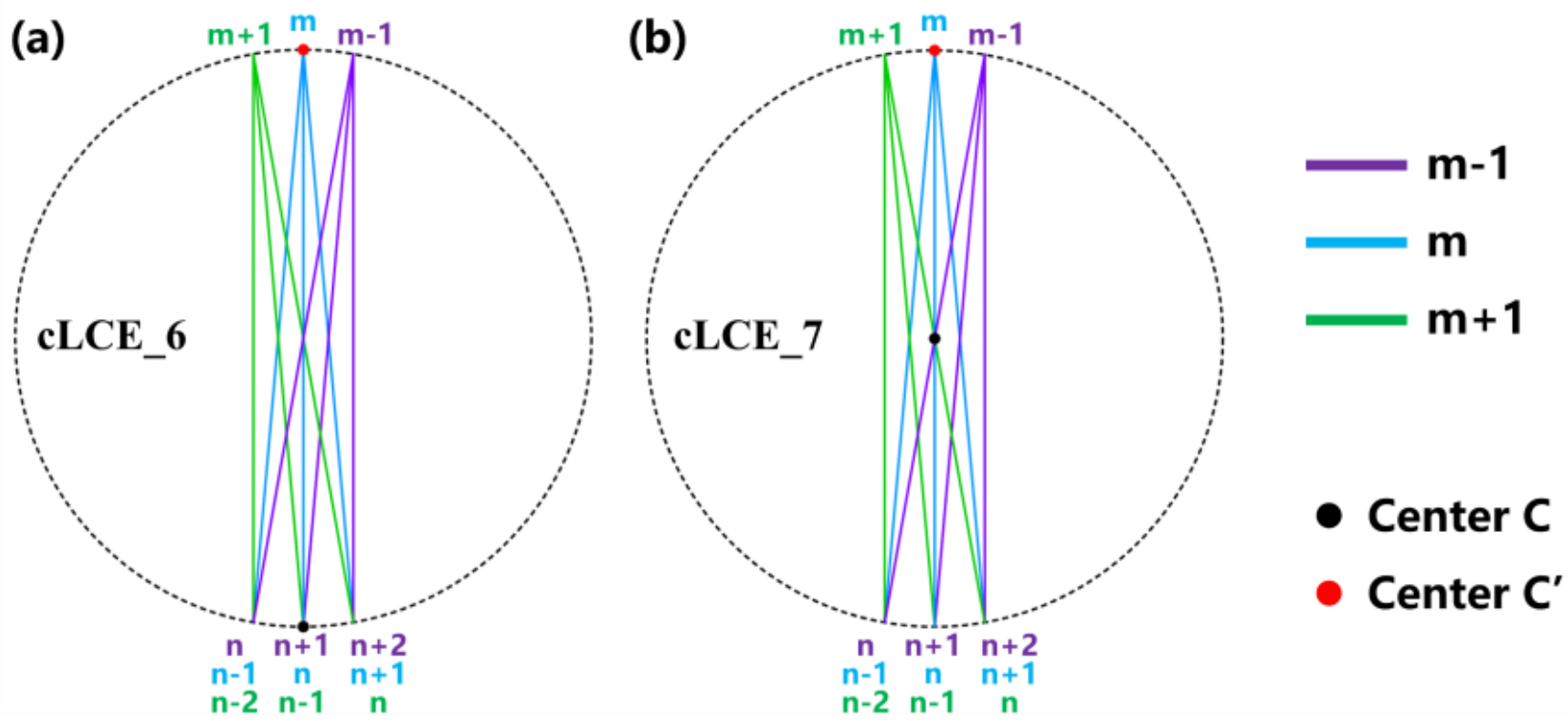

Figure 6. Discretization patterns for 2nd order cLCEs

Based on the scheme shown in Figure 6, Equation (6) can be discretized as follows:

$$\frac{\partial^2 R}{\partial t^2} = \frac{R_{m-1,n+2} - 2R_{m,n} + R_{m+1,n-2}}{\Delta\omega^2 L^2}$$

$$\frac{\partial^2 R}{\partial \omega^2} = \frac{R_{m,n+1} - 2R_{m,n} + R_{m,n-1}}{\Delta\omega^2}$$

$$\frac{\partial^2 R}{\partial \nu^2} = \frac{R_{m+1,n-1} - 2R_{m,n} + R_{m-1,n+1}}{\Delta\omega^2}$$

$$\frac{\partial^2 R}{\partial \omega \partial \nu} = \left(\frac{R_{m+1,n} - R_{m+1,n-2}}{2\Delta\omega} - \frac{R_{m-1,n+2} - R_{m-1,n}}{2\Delta\omega}\right)\frac{1}{2\Delta\omega} \quad (13)$$

Substituting the discrete expressions into Equation (6), cLCE_6 can be obtained as:

**cLCE_6:**

$$R_{m+1,n-2} - 2R_{m,n} + R_{m-1,n+2} = \left(R_{m,n+1} - 2R_{m,n} + R_{m,n-1}\right) + \left(R_{m+1,n-1} - 2R_{m,n} + R_{m-1,n+1}\right) - \frac{1}{2}\left(R_{m+1,n} - R_{m+1,n-2} - R_{m-1,n+2} + R_{m-1,n}\right) \quad (14)$$

Similarly, Equation (4) can be discretized using the same finite difference framework, yielding:

$$\frac{\partial^2 R}{\partial t^2} = \frac{R_{m-1,n+2} - 2R_{m,n} + R_{m+1,n-2}}{\Delta\omega^2 L^2}$$

$$\frac{\partial^2 R}{\partial \omega^2} = \frac{R_{m,n+1} - 2R_{m,n} + R_{m,n-1}}{\Delta\omega^2}$$

$$\frac{\partial^2 R}{\partial \theta^2} = \left(\frac{R_{m+1,n} - R_{m,n}}{2\Delta\omega} - \frac{R_{m,n} - R_{m-1,n}}{2\Delta\omega}\right)\frac{1}{2\Delta\omega}$$

$$\frac{\partial^2 R}{\partial \theta \partial t} = \left(\frac{R_{m,n+1} - R_{m-1,n+1}}{2\Delta\omega} - \frac{R_{m+1,n-1} - R_{m,n-1}}{2\Delta\omega}\right)\frac{1}{\Delta\omega L} \quad (15)$$

Based on Equation (4) and its discretized form, cLCE_7 is obtained:

**cLCE_7:**

$$R_{m+1,n-2} - 2R_{m,n} + R_{m-1,n+2} = 4\left(R_{m,n+1} - 2R_{m,n} + R_{m,n-1}\right) - \left(R_{m+1,n} - 2R_{m,n} + R_{m-1,n}\right) - 2\left(R_{m,n+1} - R_{m-1,n+1} - R_{m+1,n-1} + R_{m,n-1}\right) \quad (16)$$

In addition to the seven cLCEs directly derived from the continuous LCE formulation, two additional cLCEs can be constructed by algebraically combining pairs of the previously derived equations. Specifically, cLCE_8 is obtained by combining cLCE_1 and cLCE_2, in which the term $\frac{\Delta R}{\Delta t}$ is eliminated:

**cLCE_8:**

$$R_{m+1,n} - R_{m-1,n} = R_{m,n+1} - R_{m,n-1} + (R_{m+1,n-1} - R_{m-1,n+1}) \quad (17)$$

Similarly, cLCE_9 is derived by combining cLCE_6 and an altered form of cLCE_7, leading to the elimination of the 2nd order term $\frac{\partial^2 R}{\partial t^2}$:

**cLCE_9:**

$$R_{m,n+1} - 2R_{m,n} + R_{m,n-1} = \frac{R_{m+1,n+1} - 2R_{m+1,n} + R_{m+1,n-1}}{2} + \frac{R_{m-1,n+1} - 2R_{m-1,n} + R_{m-1,n-1}}{2} \quad (18)$$

Taken together, a total of nine discretized cLCEs are established for clinical CT geometry, including five 1st order equations (cLCE_1-cLCE_5), two 2nd order equations (cLCE_6 and cLCE_7), and two additional equations (cLCE_8 and cLCE_9) constructed through combinations of the original formulations.

## 4. Numerical verification of cLCE and Sparse-view sinogram interpolation

In this section, we numerically validate the proposed discretized cLCEs using projection data generated under a clinical CT geometry, and further demonstrate their effectiveness in recovering complete-view sinograms from sparse-view measurements for image reconstruction.

Before presenting the experimental results, several issues related to the circular scanning configuration need to be clarified. In the cLCE discretization scheme, $M$ views per rotation and $N$ detector elements per view are assumed, where $\Delta\omega$ denotes both the angular interval between adjacent views along the circular trajectory and the angular spacing between adjacent rays emitted from the same source position. This assumption leads to a uniform angular sampling framework that simplifies the derivation and implementation of the discretized cLCEs.

However, in practical clinical CT systems, the actual number of acquired views per rotation, denoted as $M'$, is significantly smaller than $M$, and the ratio $M/M'$ is typically greater than 2 slightly. This discrepancy arises because the cLCE formulation requires a denser angular sampling to fully exploit the local correlation structure of the Radon transform, whereas clinical systems operate with a more limited number of projections sufficient for standard reconstruction. In this study, $M$ is assumed to be an even number. In the special case where $M/2$ is odd, an equivalent set of projection data corresponding to the $M$-view configuration can be obtained using only $M/2$ views. This is because, under the $M$-view configuration, each beam path is sampled twice due to angular symmetry, whereas in the corresponding $M/2$-view configuration, each beam path is sampled only once.

To distinguish between these two configurations, we define the projection dataset with $M$ views per rotation as the virtual complete view (VCV), and the dataset with $M'$ views per rotation as the clinical complete view (CCV). The VCV represents an over-sampled projection domain used for theoretical analysis and numerical verification, while the CCV corresponds to the physically acquired data in clinical CT systems. In the following experiments, the ratio $M/M'$is fixed to 2

Table 1. Clinical CT geometry and acquisition parameters used in the experiments

| Parameters | | Value |
|---|---|---|
| Image size | | 512×512 pixels |
| Reconstruction grid size | | 512×512 pixels |
| Pixel size | | 0.82 mm |
| Source-to-rotation-center distance | | 595 mm |
| Detector number | | 689 |
| Fan angle | | [-26.6°, 26.6°] |
| Number of views per rotation | VCV | 2328 |
| | CCV | 1164 |
| | 1/2 sparse-view | 582 |
| | 1/4 sparse-view | 291 |
| | 1/6 sparse-view | 194 |

The circular scanning configuration used in this study, based on a clinical CT system, is summarized in Table 1. A chest CT scan (Figure 7 (a)) and an abdominal CT scan (Figure 7 (b)) from Mayo Low Dose challenge (www.aapm.org/GrandChallenge/LowDoseCT/) are used in both the numerical verification and the sparse-view sinogram interpolation experiments.

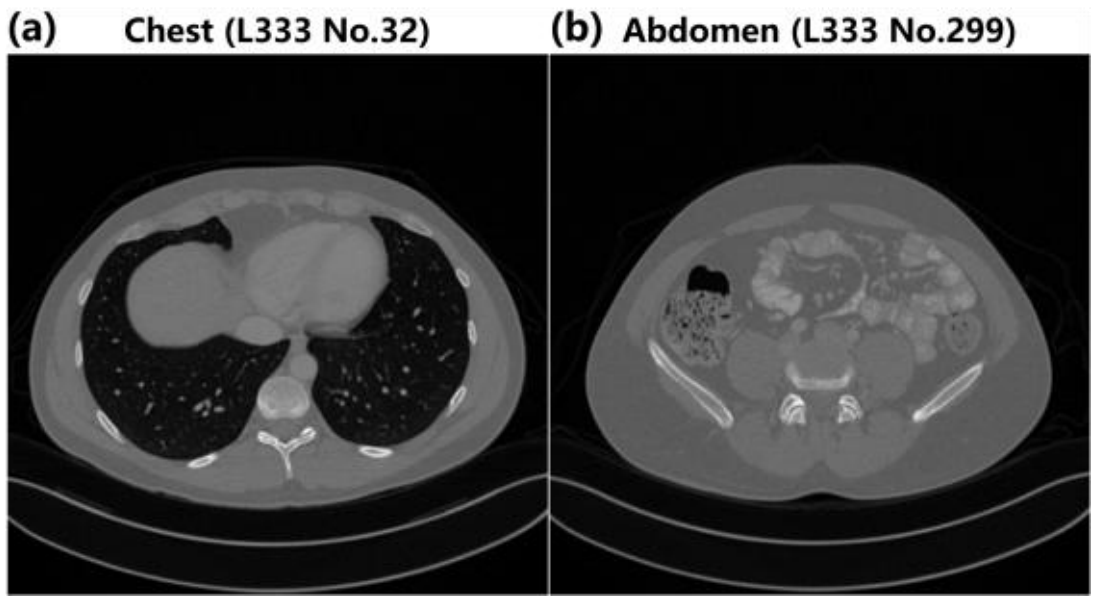

Figure 7. Clinical CT datasets (chest and abdomen) used in experiments (displayed in full intensity range)

### 4.1 Numerical verification of cLCE

The numerical verification of the proposed cLCEs is performed using projection data generated under the VCV configuration. It should be noted that, when evaluating the discretized cLCEs using projection data ($\hat{R}$) generated from discrete images, both discretization errors and quantization errors are inevitably introduced. As a result, the equality implied by the continuous formulation cannot be satisfied exactly in practice. To quantify this discrepancy, we define the cLCE residual $\varepsilon_i(\hat{R})$ for the i-th equation ($i = 1,\cdots,9$) as the difference between the left-hand side and the right-hand side of the corresponding discretized cLCE:

$$\varepsilon_i(\hat{R}) = cLCE_i(\hat{R}) = LHS_i(\hat{R}) - RHS_i(\hat{R}) \tag{19}$$

where $\hat{R}$ denotes the projection data used for evaluation, $LHS_i$ for the left-hand side of the i-th cLCE, and $RHS_i$ for the right-hand side of the i-th cLCE. Take cLCE_6 for example, the residual of cLCE_6 evaluated at $\hat{R}_{m,n}$ can be written as:

$$\begin{aligned}\varepsilon_6(\hat{R}_{m,n}) = \hat{R}_{m+1,n-2} - 2\hat{R}_{m,n} + \hat{R}_{m-1,n+2} - (\hat{R}_{m,n+1} - 2\hat{R}_{m,n} + \hat{R}_{m,n-1}) \\ -(\hat{R}_{m+1,n-1} - 2\hat{R}_{m,n} + \hat{R}_{m-1,n+1}) \\ +\frac{1}{2}(\hat{R}_{m+1,n} - \hat{R}_{m+1,n-2} - \hat{R}_{m-1,n+2} + \hat{R}_{m-1,n})\end{aligned} \tag{20}$$

To further quantify the accuracy, two evaluation metrics are employed.

(1) The mean absolute error (MAE) is computed as the average absolute value of the cLCE residual:

$$MAE_i = \frac{\sum_{m=0}^{M-1}\sum_{n=0}^{N-1}|\varepsilon_i(\hat{R}_{m,n})|}{M\cdot N} \qquad (i = 1,\cdots,9) \tag{21}$$

(2) The mean percentage error (MPE) is defined as the average ratio between the absolute residual and the corresponding central projection value $\hat{R}_{m,n}$, where $\hat{R}_{m,n}$ denotes the projection sample at which the cLCE is evaluated:

$$MPE_i = \frac{100}{M\cdot N}\cdot\sum_{m=0}^{M-1}\sum_{n=0}^{N-1}\frac{|\varepsilon_i(\hat{R}_{m,n})|}{\hat{R}_{m,n}} \qquad (i = 1,\cdots,9) \tag{22}$$

During the numerical verification of the discretized cLCEs, special care is taken in the computation of the derivative terms on both sides of the equations. Specifically, for cLCE_1-cLCE_3, cLCE_6 and cLCE_7, the left side terms (i.e., $\frac{\Delta R}{\Delta t}$ and $\frac{\partial^2 R}{\partial t^2}$) are kept unchanged, while a smoothing filter is applied to the derivative terms on the right side, including $\frac{\Delta R}{\Delta\omega}$, $\frac{\Delta R}{\Delta\nu}$, $\frac{\Delta R}{\Delta\theta}$ and 2nd order terms. For cLCE_4, cLCE_5, cLCE_8, and cLCE_9, the right side terms remain unchanged, and a smoothing filter is applied to the left side term.

This asymmetric treatment is motivated by the following considerations. First, the projection data are generated from discretized images, and the cLCEs themselves are approximated using central difference schemes, which inherently introduce discretization errors. As a result, direct evaluation of both sides may amplify discretization-induced inconsistencies. To mitigate this issue, filtering is applied to the derivatives of only one side to suppress high-frequency numerical fluctuations and reduce discretization-induced errors. Meanwhile, the other side is intentionally left unfiltered to avoid over-smoothing, thereby preserving the fidelity of the reference term used for comparison.

The filtering is performed along the natural direction of each derivative operator. For example, consider the term $\frac{\Delta R}{\Delta \omega}$, where $\omega$ denotes the detector coordinate at a fixed source position m. In the discretized form of cLCE_1, this derivative is approximated as $(R_{m,n+1} - R_{m,n-1})$. Accordingly, the filtering is applied along the detector direction, such that each term (i.e., $R_{m,n+1}$) is replaced by a locally weighted average of neighboring detector samples. Specifically, the filtered value can be express as:

$$\tilde{R}_{m,n+1} = \sum_{i=-k}^{k} w_i R_{m,n+1+i} \tag{23}$$

where $\tilde{R}_{m,n+1}$ is the filtered projection data, $w_i$ defines a smoothing kernel along the detector direction and satisfy $\sum w_i = 1$.

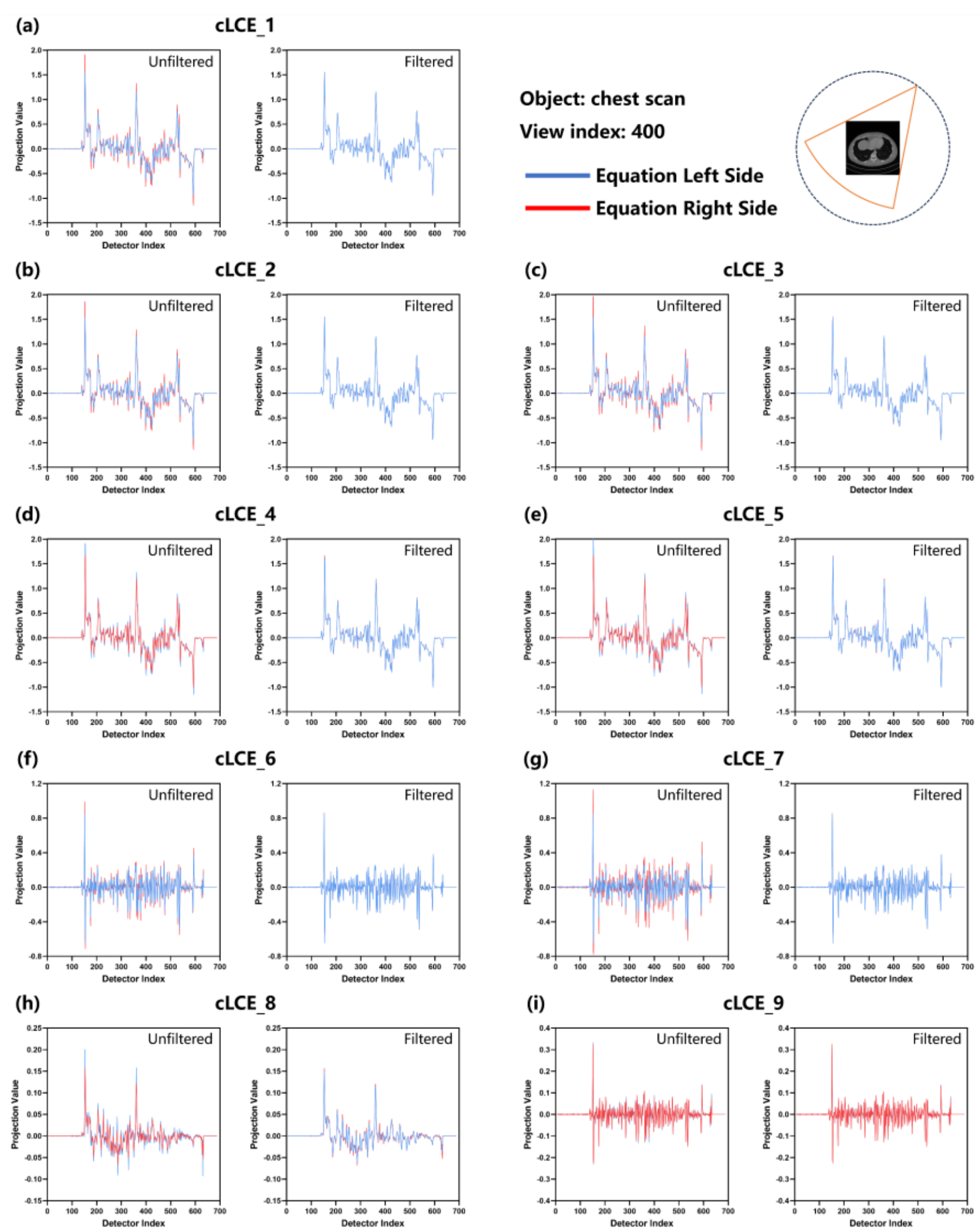

Figure 8. Numerical consistency evaluation of cLCEs using chest projection data (view index = 400)

A similar filtering strategy is applied to the other derivative terms of the cLCEs. In each case, the smoothing operation is aligned with the corresponding derivative direction to ensure consistency with the underlying discretization scheme. This approach provides a balanced trade-off between noise suppression and structural preservation, enabling a more reliable numerical verification of the proposed cLCEs.

The consistency examination of the nine proposed cLCE based on projection data generated from clinical chest scan and abdominal scan are presented in Figure 8 and Figure 9, respectively. Both unfiltered and filtered data with the view index of 400 are shown. Error introduced by discretization errors are noticeable under the unfiltered condition, while the two sides of the equation converge well under the filtered condition.

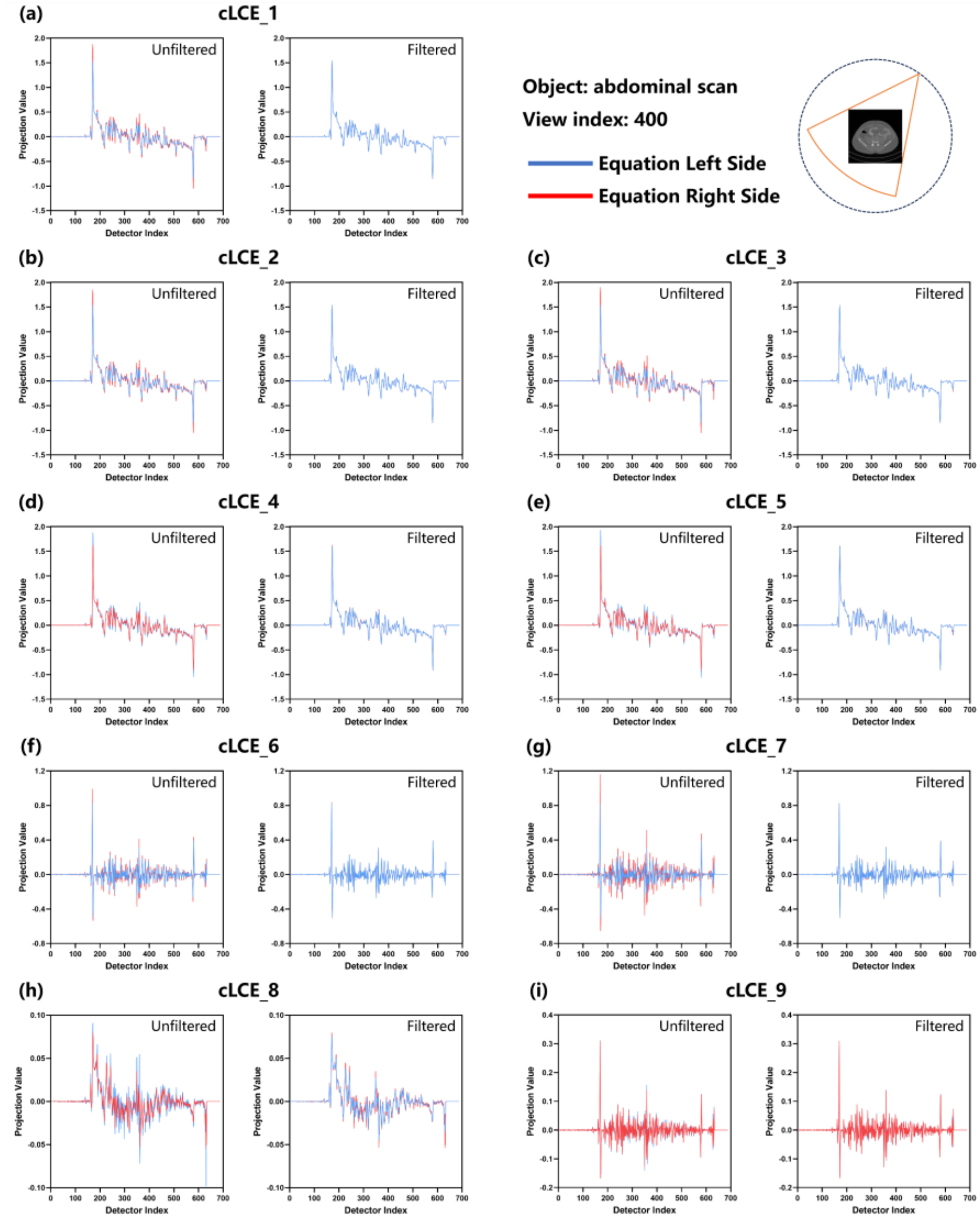

Figure 9. Numerical consistency evaluation of cLCEs using abdominal projection data (view index = 400)

The MAE and MPE of the cLCE residuals, computed using both unfiltered and filtered data, are summarized in Table 2. It should be noted that cLCE_8 and cLCE_9 involve different left side terms compared to the other seven cLCEs. As a result, their corresponding MAE and MPE values are not directly comparable with those of cLCE_1-cLCE_7. Although cLCE_8 and

cLCE_9 yield smaller numerical errors, this improvement is primarily due to the altered formulation rather than a direct enhancement in numerical consistency. Therefore, for a fair comparison among equations with consistent formulations, cLCE_6 is identified as exhibiting the best performance, achieving the smallest MAE under both unfiltered and filtered conditions. Based on this observation, cLCE_6 is selected for the subsequent sparse-view sinogram interpolation experiments.

Table 2. Quantitative evaluation of cLCE consistency (MAE and MPE)

| Object | cLCE | Unfiltered | | Filtered | |
|---|---|---|---|---|---|
| | | MAE | MPE | MAE | MPE |
| Chest scan | cLCE_1 | 0.0309 | 3.79% | 0.0030 | 0.94% |
| | cLCE_2 | 0.0313 | 3.82% | 0.0033 | 1.10% |
| | cLCE_3 | 0.0313 | 3.82% | 0.0033 | 1.10% |
| | cLCE_4 | 0.0215 | 2.98% | 0.0062 | 1.04% |
| | cLCE_5 | 0.0215 | 2.98% | 0.0062 | 1.04% |
| | cLCE_6 | 0.0131 | 2.83% | 0.0013 | 0.38% |
| | cLCE_7 | 0.0262 | 3.80% | 0.0019 | 0.43% |
| | cLCE_8 | 0.0055 | 1.34% | 0.0019 | 0.64% |
| | cLCE_9 | 0.0023 | 0.65% | 0.0019 | 0.57% |
| Abdomen scan | cLCE_1 | 0.0253 | 3.42% | 0.0029 | 0.75% |
| | cLCE_2 | 0.0258 | 3.46% | 0.0032 | 0.92% |
| | cLCE_3 | 0.0258 | 3.46% | 0.0032 | 0.92% |
| | cLCE_4 | 0.0175 | 2.56% | 0.0048 | 0.86% |
| | cLCE_5 | 0.0175 | 2.56% | 0.0048 | 0.86% |
| | cLCE_6 | 0.0117 | 2.46% | 0.0013 | 0.33% |
| | cLCE_7 | 0.0233 | 3.72% | 0.0019 | 0.40% |
| | cLCE_8 | 0.0049 | 1.12% | 0.0018 | 0.54% |
| | cLCE_9 | 0.0023 | 0.57% | 0.0019 | 0.50% |

### 4.2 Sparse-view sinogram interpolation

At the beginning of this section, it is worth clarifying the notation used for projection data. In Section 4.1, the symbol $\hat{R}$ is introduced to denote the projection data generated from discrete images for the purpose of numerical consistency evaluation. In the present section, however, we revert to the notation $R$ to represent the projection data used in the sinogram interpolation experiments, in order to maintain consistency with the formulation of the optimization problem and avoid notational ambiguity.

The sparse-view sinogram interpolation is implemented through the following procedure.

First, sparse-view projection data are generated by downsampling the CCV projections according to a given sparsity ratio (the sparsity level is defined with respect to the CCV configuration, refer to Table 1).

Second, a preliminary estimate of the VCV projection data is obtained by applying conventional interpolation methods, such as linear interpolation or spline interpolation, to the sparse-view data.

Third, the proposed cLCE-based method is employed to refine the interpolated projections. Among the discretized equations, cLCE_6 is selected due to its superior performance in the numerical verification experiments, and both the unfiltered and filtered versions of cLCE_6 are evaluated to investigate their effectiveness in refining the initial VCV estimate. Let $\boldsymbol{R}$ ($\boldsymbol{R} \subseteq R^{MN\times1}$) denote the vectorized interpolated VCV sinogram, originally of size $M \times N$, where $M$ and $N$ correspond to the number of views and detector elements, respectively. The refinement problem is formulated as the following optimization problem:

$$\min_{\boldsymbol{R}}\{\|\boldsymbol{PR}\|_2^2 + \alpha\|\boldsymbol{DR} - \boldsymbol{R_s}\|_2^2\} \tag{24}$$

where the first term enforces consistency with the cLCE constraint, and the second term preserves agreement with the known sparse-view measurements. Here, $\boldsymbol{P}$ ($\boldsymbol{P} \subseteq R^{MN\times MN}$) is a system matrix constructed from the discretized cLCE coefficients, where each row corresponds to a local cLCE constraint centered at a specific projection sample in $\boldsymbol{R}$. The matrix $\boldsymbol{D}$ ($\boldsymbol{D} \subseteq R^{MN\times MN}$) is a diagonal mask matrix that selects the locations of the known sparse-view projections, and $\boldsymbol{R_s}$ ($\boldsymbol{R_s} \subseteq R^{MN\times1}$) denotes the sparse-view projection data, with unmeasured entries set to zero. The parameter $\alpha$ is a relaxation factor that balances the cLCE consistency term and the data fidelity term.

This formulation corresponds to a quadratic inverse problem, for which various regularized optimization methods, such as ridge regression (Hoerl and Kennard 1970) or Lasso (Santosa and Symes 1986), could be employed. However, to evaluate the intrinsic capability of the proposed cLCE without introducing additional prior information, we adopt a gradient-descent-based (Landweber 1951) optimization scheme without explicit regularization. This design choice ensures that the observed improvements can be primarily attributed to the cLCE formulation itself, rather than to regularization effects.

Finally, filtered backprojection (FBP) is used to reconstruct images from sparse-view data, CCV data, and the cLCE-refined VCV data. The reconstruction results are then compared to assess the effectiveness of the proposed cLCE-based interpolation method. To quantitatively evaluate the reconstruction quality, peak signal-to-noise ratio (PSNR) and structural similarity index (SSIM) are employed as evaluation metrics. PSNR and SSIM are formulated as below:

$$PSNR = 10 \cdot log_{10}\left(\frac{MAX^2}{MSE}\right) \tag{25}$$

In Equation (25), MAX denote the maximum possible pixel value and MSE is the mean squared error between the reconstructed image and the reference image.

$$SSIM(x,y) = \frac{(2\mu_x\mu_y+C_1)(2\delta_{xy}+C_2)}{({\mu_x}^2+{\mu_y}^2+C_1)({\delta_x}^2+{\delta_y}^2+C_2)} \tag{26}$$

In Equation (26), $\mu_x$ and $\mu_y$ are the mean intensity of the image, $\delta_x$ and $\delta_y$ are the variance, $\delta_{xy}$ is the covariance, $C_1$ and $C_2$ are small constants for numerical stability.

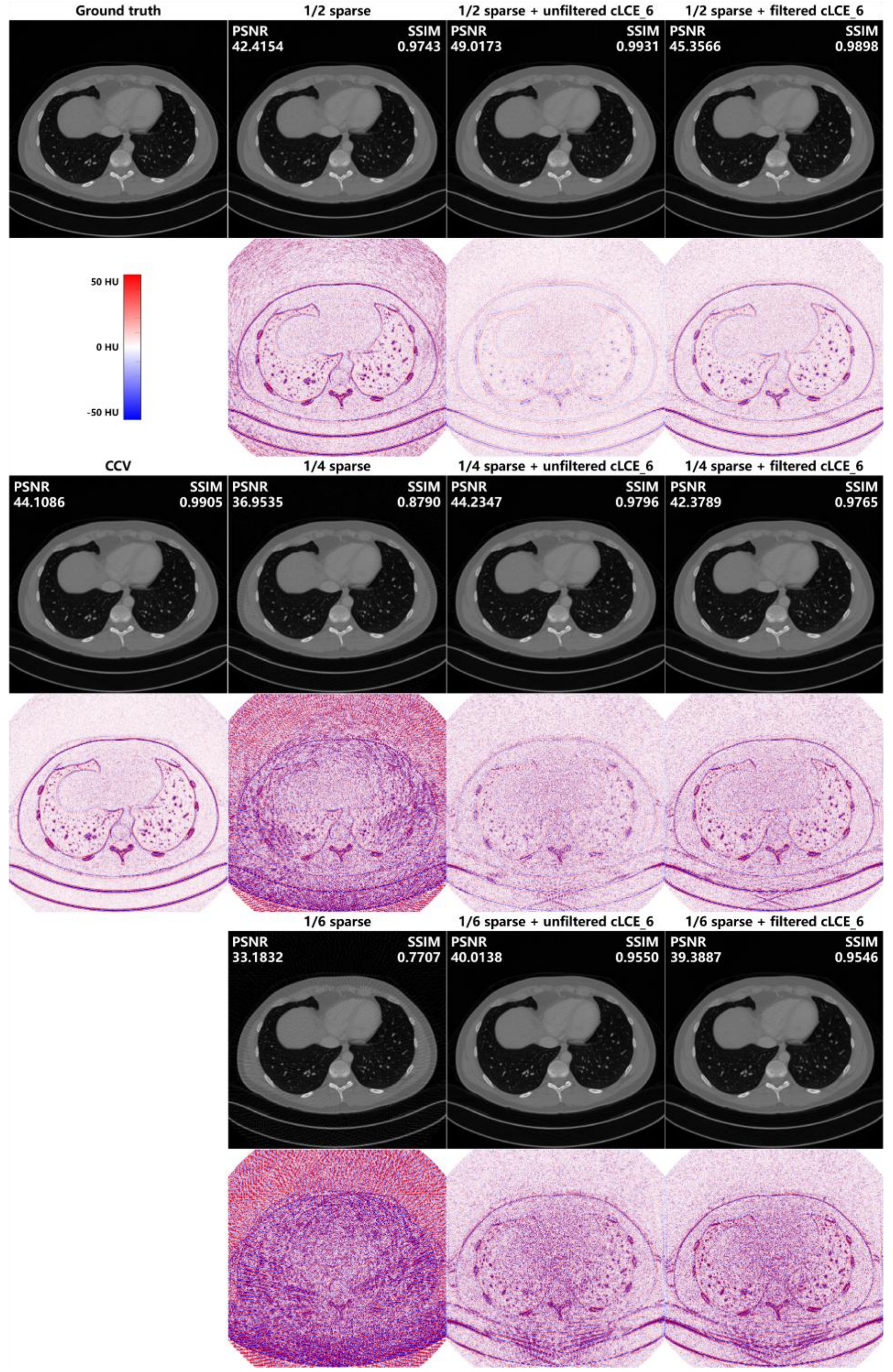

Figure 10. FBP reconstruction from sparse-view, CCV, and cLCE-refined VCV projections for the chest dataset, along with corresponding difference images

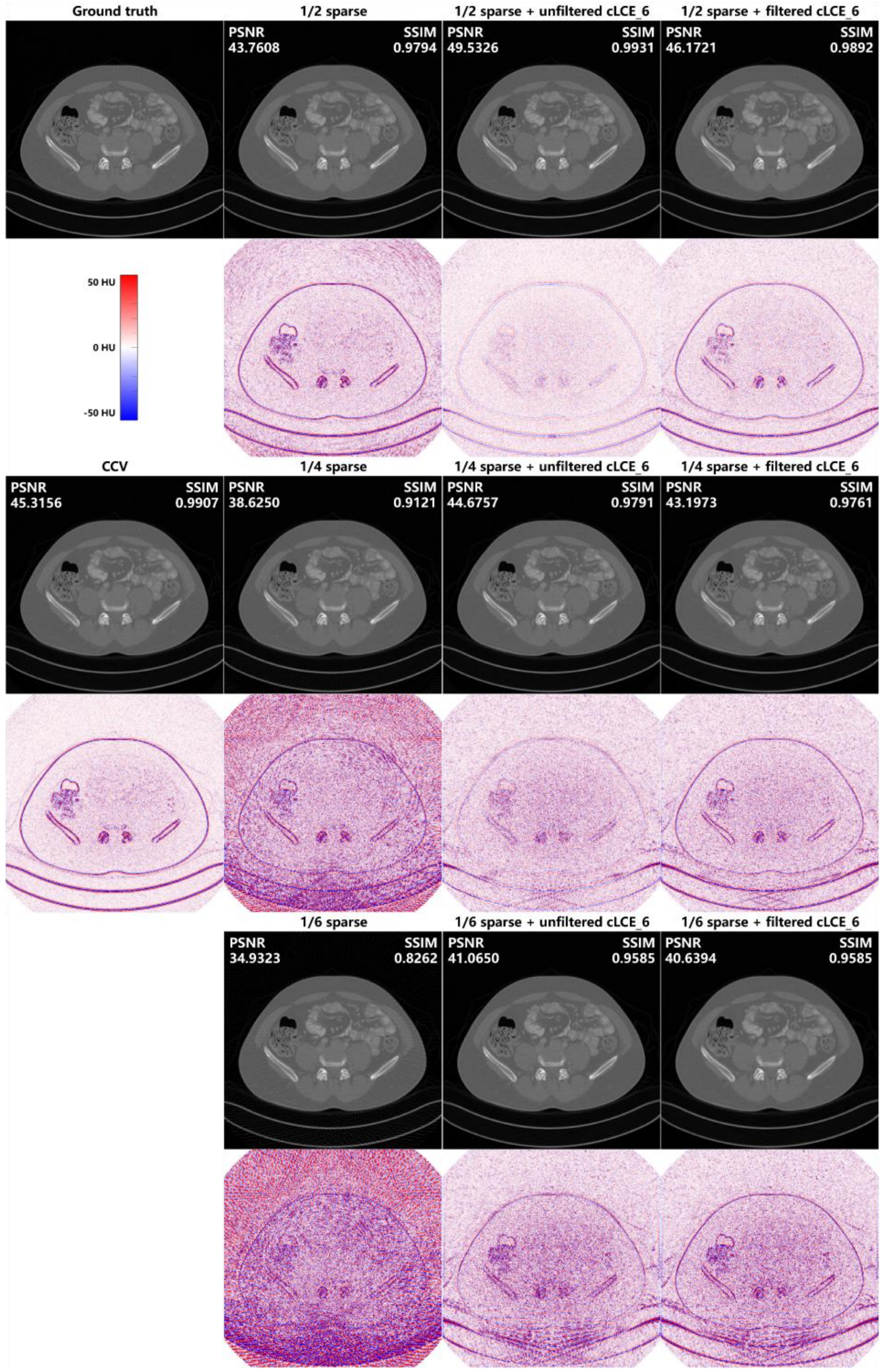

Figure 11. FBP reconstruction from sparse-view, CCV, and cLCE-refined VCV projections for the abdominal dataset, along with corresponding difference images

The FBP reconstruction result from sparse-view projection, CCV data, and cLCE-refined VCV data, as well as their difference image from the ground truth are shown in Figure 10 and Figure 11. The results show that cLCE-based refinement significantly improves reconstruction quality compared to sparse-view reconstruction, achieving approximately 6–7 dB improvement in PSNR under 1/2, 1/4, and 1/6 sparse-view conditions when using the unfiltered cLCE_6. In addition, the results refined by the unfiltered cLCE_6 outperform CCV-based reconstruction, with a PSNR gain of approximately 4-5 dB in the 1/2 sparse-view case and a marginal improvement (~0.2 dB) in the 1/4 sparse-view case. The difference image indicates that cLCE refinement strongly improves the edge structure and reduces artifacts introduced by sparse-view projection.

In contrast, the filtered cLCE_6 exhibits consistently lower performance across all tested sparsity levels. At the 1/2 sparse-view level, the filtered cLCE_6 achieves approximately 2–3 dB PSNR improvement, which is significantly lower than that of the unfiltered formulation. At the 1/4 sparse-view level, the filtered version remains approximately 1.5–2 dB lower in PSNR. As the sparsity increases to the 1/6 level, the performance gap decreases to less than 1 dB.

These results indicate that, although filtering improves numerical consistency in the cLCE verification stage, the unfiltered cLCE_6 provides superior performance in sparse-view sinogram refinement across all tested conditions. This suggests that the unfiltered cLCE formulation performs effective in recovering complete sinograms from sparse-view data, even in the presence of discretization errors. One possible explanation is that the filtering operation introduces additional smoothing, which may weaken fine structural information in the sinogram, thereby reducing reconstruction accuracy.

Overall, the numerical verification results confirm that the discretized cLCEs provide accurate approximations of the underlying PDE relationships, with cLCE_6 demonstrating the best numerical consistency. In addition, the sparse-view interpolation experiments show that the proposed cLCE-based refinement effectively improves projection quality and leads to substantial gains in reconstructed image quality. These findings validate both the numerical correctness of the discretized cLCEs and their practical applicability in sparse-view CT reconstruction.

## 5. Discussion and Conclusion

In this work, we extended the local correlation equation (LCE) framework of the 2D Radon transform to clinical CT geometry and developed a set of discretized circular LCEs (cLCEs) based on practical acquisition configurations which can be directly applied to clinical projection data.

Numerical verification results demonstrate strong consistency between the two sides of the discretized equations, particularly for the 2nd order equation cLCE_6, which exhibits the lowest residual error among all tested formulations. The introduction of directionally aligned filtering effectively mitigates discretization-induced errors while preserving the fidelity of the reference term, leading to more reliable validation results.

In sparse-view sinogram interpolation experiments, the proposed cLCE-based refinement significantly improves reconstruction quality over conventional interpolation methods. Notably, the refined projections can outperform reconstruction based on clinical complete view (CCV) data under certain sampling conditions, indicating that the cLCE framework effectively exploits the intrinsic redundancy of the Radon transform. Although directionally aligned filtering significantly reduces the residual in the numerical verification stage, comparative experiments show that the unfiltered cLCE_6 consistently achieves superior reconstruction performance over its filtered counterpart across all tested sparsity levels. Despite the presence of discretization errors, the unfiltered formulation yields higher PSNR and better structural preservation, demonstrating that the cLCE remains effective for recovering complete sinograms from sparse-view data. This observation suggests that improved numerical consistency does not necessarily translate to better reconstruction performance. A possible explanation is that the filtering operation introduces additional smoothing, which may suppress fine structural details in the sinogram. Overall, these results indicate that the cLCE formulation is inherently stable and robust when applied to practical projection data, even without explicit error compensation.

Despite these promising results, several limitations remain. The current implementation employs a simple gradient-descent scheme without additional regularization, which may limit robustness under high noise conditions. Furthermore, only a subset of discretized cLCEs is explored in the interpolation task, and more advanced combinations or optimization strategies could further enhance performance. Future work will focus on integrating the cLCE framework with regularized reconstruction models and exploring its application in ultra-sparse-view CT imaging scenarios.

## Data availability statement

All data that support the findings of this study are included within the article.

## Acknowledgment

This work was supported by the National Key Research and Development Program of China (Nos.2022YFA1204203, 2016YFA0202003). We would like to thank Professor Zhiwei Qiao Shanxi University and Professor Shaojie Tang with Xi'an University of Posts & Telecommunications for the helpful discussions on our study and their invaluable advice. We also would like to thank Dr Cynthia McCollough, the Mayo Clinic, the American Association of Physicists in Medicine, and the National Institute of Biomedical Imaging and Bioengineering for providing the Mayo Clinic data.